\documentclass[12pt,preprint]{aastex}

\newcommand{\ldot}{L$_{\odot}$}
\newcommand{ \um}{$\mu$m~}
\newcommand{ \ums}{$\mu$m}
\def\kmsMpc{\ifmmode {\rm\,km\,s^{-1}\,Mpc^{-1}}\else
    ${\rm\,km\,s^{-1}\,Mpc^{-1}}$\fi}

\shorttitle{Luminosities of Dusty Quasars}
\shortauthors{Vardanyan et al.}

\begin{document}

\title{Seeking the Epoch of Maximum Luminosity for Dusty Quasars }

\author{Valeri Vardanyan\altaffilmark{1,2,3}, Daniel Weedman\altaffilmark{4}, and Lusine Sargsyan\altaffilmark{4}} 

\altaffiltext{1}{Department of Theoretical Physics, Faculty of Physics, Yerevan State University, Yerevan, 0025, Armenia; vardanyanv@gmail.com}
\altaffiltext{2}{Theory Division, Alikhanian National Laboratory (Yerevan Physics Institute), Yerevan, 0036, Armenia}
\altaffiltext{3}{Fakult{\"a}t f{\"u}r Physik und Astronomie, Ruprecht-Karls-Universit{\"a}t Heidelberg, Heidelberg, Germany}
\altaffiltext{4}{Center for Radiophysics and Space Research, Cornell University, Ithaca,
NY 14853, USA; dweedman@isc.astro.cornell.edu}

\begin{abstract}

Infrared luminosities $\nu L_{\nu}$(7.8 \ums) arising from dust reradiation are determined for Sloan Digital Sky Survey (SDSS) quasars with 1.4 $<$ z $<$ 5 using detections at 22 \um by the Wide-Field Infrared Survey Explorer.  Infrared luminosity does not show a maximum at any redshift z $<$ 5, reaching a plateau for z $\ga$ 3 with maximum luminosity $\nu L_{\nu}$(7.8 \ums) $\ga$ 10$^{47}$ erg s$^{-1}$; luminosity functions show one quasar Gpc$^{-3}$ having $\nu L_{\nu}$(7.8 \ums) $>$ 10$^{46.6}$ erg s$^{-1}$ for all 2 $<$ z $<$ 5.  We conclude that the epoch when quasars first reached their maximum luminosity has not yet been identified at any redshift below 5.  The most ultraviolet luminous quasars, defined by rest frame $\nu L_{\nu}$(0.25 \ums), have the largest values of the ratio $\nu L_{\nu}$(0.25 \ums)/$\nu L_{\nu}$(7.8 \ums) with a maximum ratio at z = 2.9.  From these results, we conclude that the quasars most luminous in the ultraviolet have the smallest dust content and appear luminous primarily because of lessened extinction.  Observed ultraviolet/infrared luminosity ratios are used to define ``obscured" quasars as those having $>$ 5 magnitudes of ultraviolet extinction.  We present a new summary of obscured quasars discovered with the $Spitzer$ Infrared Spectrograph and determine the infrared luminosity function of these obscured quasars at z $\sim$ 2.1.  This is compared with infrared luminosity functions of optically discovered, unobscured quasars in the SDSS and in the AGN and Galaxy Evolution Survey. The comparison indicates comparable numbers of obscured and unobscured quasars at z $\sim$ 2.1 with a possible excess of obscured quasars at fainter luminosities.  

\end{abstract}


\section{Introduction}

Understanding the formation and evolution of the most massive galaxies and the most massive black holes powering active galactic nuclei (AGN) is a fundamental goal for observational and theoretical cosmology.  When did these objects form? What processes assembled them?  What is the relation between star formation and black hole formation?  Seeking the most luminous, most extreme quasars is a straightforward observational goal to determine some of these answers.  These quasars can be traced to the highest redshifts, and the most luminous, most massive galaxies and quasars are those whose formation is most difficult to explain.  The challenges in understanding the formation mechanisms and formation epoch for the most massive galaxies are recently summarized by \citet{tof14} and for the earliest supermassive black holes by \citet{fen14}.

Evidence that the formation processes of stars, galaxies and AGN are related arises from the observed correlation between concentrated masses in the centers of galaxies (presumed to be supermassive black holes) and the mass of the surrounding halo of stars \citep[e.g.][]{fer00,geb00}.  As a result, preferred scenarios form large numbers of stars rapidly in extended haloes while simultaneously feeding and enlarging the central black holes, thereby powering a central AGN \citep[e.g.][]{hop06, hop08}.  Scenarios which connect the formation of massive galaxies and supermassive black holes imply that finding the most luminous quasars in the early universe also determines the epoch when the most massive galaxies were forming.

Such scenarios have been pursued in increasing detail for 25 years, beginning with an attempt \citep{san88} to explain the Ultraluminous Infrared Galaxies (ULIRGs). The explanation proposed for these very dusty, very luminous sources was that they begin with the merger of smaller gas rich galaxies. The merging process allows sufficient angular momentum loss that large amounts of gas can settle into dense clouds which produce extensive starbursts while some gas continues to the center to power accretion and enlarge any existing black hole \citep{bar96}. Dust produced by the rapid evolution of massive stars obscures the AGN and reradiates in the infrared the large intrinsic luminosity produced from the core regions at shorter wavelengths. As the radiated power of the accretion disk increases, the surrounding dust is blown away, and an unobscured central quasar is revealed \citep{hop06,nar10}.

Observations of the irregular, merged structure of ULIRGs \citep{vei95}, of the changing infrared spectral classification of starbursts to AGN during the evolving process \citep{far09}, and of the required winds for gas ejection \citep{stu11,rup13} strengthen this scenario observationally.  Modeling \citep{nar10} predicts the timeline and luminosity behavior of the dusty merger process which leads to both a massive galaxy and a luminous quasar.


Existing models match long standing observations of quasar evolution which showed that the peak luminosity of optically discovered quasars occurs at a redshift z $\ga$ 2.  Observational development of this conclusion began with the demonstration by \citet{sch68} that quasars were systematically more luminous in the past and was refined through various quasar surveys \citep{car78,lew79,osm82,sch83,mar84,boy88,fan04,cro04}.  This epoch of maximum luminosity for quasars is similar to the epoch initially determined as that of maximum star formation \citep[e.g.][]{mad98, red09}, which encouraged the development of scenarios to explain how both events can occur together. 

Recent observations of dusty sources not present in optical surveys suggest, however, that very luminous and rapid star formation together with the associated quasars occurred much earlier, at redshifts exceeding 4 \citep{cas12,swi12,wan13,car13,rie13}.  This is potentially an important result if it pushes the formation epoch of the most massive galaxies and most luminous quasars to an earlier time; the difference between z = 2 and z = 4 halves the time available for massive galaxy and black hole formation (down to 1.5 Gyr from the beginning of the Universe at z = 4 from 3.2 Gyr at z = 2). 

Because of observations and models which indicate that the earliest phases of luminous quasars and starbursts may be characteristically dusty, it is important to trace luminous sources using their dust luminosity because the dust can diminish or extinguish the optical and ultraviolet luminosity.  We pursue several objectives in the present paper by using dust luminosity to trace the most luminous quasars as a function of redshift.  Our first goal is to determine if an epoch of peak dust luminosity can be identified.  Then, we compare infrared and ultraviolet luminosities to seek any observable changes in dust content as a function of epoch or luminosity for the most luminous quasars.  Finally, we compare luminosity functions for dust-obscured and unobscured quasars to compare luminosities and to determine what fraction of luminous quasars are obscured by dust.

As a measure of dust luminosity, $\nu L_{\nu}$(7.8~\ums) (subsequently called ``infrared luminosity") is used for the reasons summarized in \citet{wee12}, who show that this correlates well with other measures of AGN luminosity such as hard X-ray luminosity, black hole mass, and high ionization emission line luminosity.  These authors also review the scaling between $\nu L_{\nu}$(7.8~\ums) and total infrared luminosity or total bolometric luminosity. The choice of an exact 7.8 \um wavelength is determined by the mid infrared continuum maximum between absorbing features in heavily obscured sources (described and illustrated below in section 4.2).  

In the present paper, the 7.8 \um luminosity is determined for several samples of quasars using newly available infrared data and used to consider three specific and related objectives.  Our first objective (Section 2) is to determine $\nu L_{\nu}$(7.8~\ums) for the brightest optically discovered quasars, which are those quasars in the quasar catalog of \citet{sch10} arising from the Sloan Digital Sky Survey (SDSS, Gunn et al. 1998). The $\nu L_{\nu}$(7.8 $\mu$m) are measured for SDSS quasars using observed 22 \um flux densities from the Wide-Field Infrared Survey Explorer (WISE, Wright et al. 2010), transformed to rest frame 7.8 \um using an empirical spectral template determined for some SDSS quasars with the $Spitzer$ Infrared Spectrograph (IRS; Houck et al. 2004). Our purpose in this analysis is to seek the epoch of maximum dust luminosity for the most luminous, unobscured quasars. 

The next objective (Section 3) is to determine if the fractional dust content of SDSS quasars changes systematically with luminosity or redshift, by comparing infrared dust luminosities $\nu L_{\nu}$(7.8 $\mu$m) with observed ultraviolet luminosities $\nu L_{\nu}$(0.25 \ums) for SDSS quasars as determined by \citet{she11}.  This comparison also leads to our quantitative definition of ``unobscured" quasars, defined by the ultraviolet to infrared luminosity ratios for SDSS quasars.  By comparing the ultraviolet/infrared luminosity ratio to redshift and ultraviolet luminosity, we test in this section some expectations of the dusty merger scenario.  

Finally, in Section 4, we compare optically discovered, unobscured quasars with a sample of obscured quasars too faint for inclusion in systematic optical surveys (Vega magnitudes $I$ $>$ 24).  Our goal in this comparison is to determine if obscured quasars have different dust luminosities than unobscured quasars, and to determine the relative numbers of each category. For this purpose, we define ``obscured" quasars as those with observed ultraviolet to infrared ratios at least 100 times less than for any SDSS quasar. These obscured quasars were discovered \citep[e.g.][]{hou05,yan07,saj07,bus09} in infrared surveys at 24 \um with the $Spitzer$ Space Telescope and have redshifts determined from the characteristic silicate absorption feature at rest frame 9.7 \um seen with the IRS.  Such sources have previously been defined as ``dust obscured galaxies" (DOGS; Dey et al. 2008).  We present a new summary of spectroscopic measurements for 37 quasars among the DOGs having z $>$ 1.4 that arise from a sample defined only by infrared and optical fluxes, $f_{\nu}$(24 \ums) $>$ 1 mJy and $I$ $>$ 24.  

This sample of obscured quasars is used to determine luminosity functions in $\nu L_{\nu}$(7.8 $\mu$m) for comparison to the luminosity functions for optically discovered quasars.  This comparison is done for 1.8 $<$ z $<$ 2.4, which is the redshift range with the most complete sample of obscured quasars.  For this comparison, the SDSS quasars are supplemented by the fainter quasars in a deeper optical survey, the AGN and Galaxy Evolution Survey (AGES; Kochanek et al. 2012), which reaches approximately a factor of 10 fainter than SDSS in both optical and infrared fluxes.  The $\nu L_{\nu}$(7.8 $\mu$m) are measured for AGES quasars using 24 \um flux densities available from $Spitzer$ photometry, yielding a luminosity function based on dust luminosities for optically discovered quasars that encompasses a factor of 400 in luminosity at z $\sim$ 2.1.  We determine luminosities using H$_0$ = 74 \kmsMpc \citep{rie11}, $\Omega_{M}$=0.27, and $\Omega_{\Lambda}$=0.73.  

These three analyses of luminous, dusty quasars produce observational results that determine whether the most luminous quasars in the universe have yet been identified, whether there is evidence for changing dust content as a function of redshift, and whether unobscured or obscured quasars dominate in the early universe.

\section{Infrared Luminosities of SDSS Quasars}

In this section, we determine the $\nu L_{\nu}$(7.8~\ums) dust luminosities of optically discovered quasars in the SDSS.  Our objective is to learn how maximum luminosities change with redshift, as determined both from the redshift distribution of observed luminosities and from an analysis of the bright end of the luminosity functions for 1.4 $<$ z $<$ 5.

\subsection{Determining Infrared Luminosities using WISE}

The optically discovered quasar sample representing the brightest known quasars is the SDSS DR7 quasar catalog of \citet{sch10}.  The SDSS quasars which we use are those uniformly selected according to photometric criteria described in \citet{ric02}; these quasars are flagged "1" in the DR7 compilation of \citet{she11}.  These SDSS quasars were compared with $f_{\nu}$(22 \ums) from the all sky WISE Source Catalog\footnote{http://wise2.ipac.caltech.edu/docs/release/allsky/}. A source identification was taken to be correct if the WISE coordinate and SDSS coordinate agree to within 3$\arcsec$.  The WISE fluxes are derived from 22 \um magnitudes given in the WISE catalog by assuming a magnitude zero point of 8.28 Jy (Vega magnitudes).

\begin{figure}

\figurenum{1}
\includegraphics[scale= 0.5]{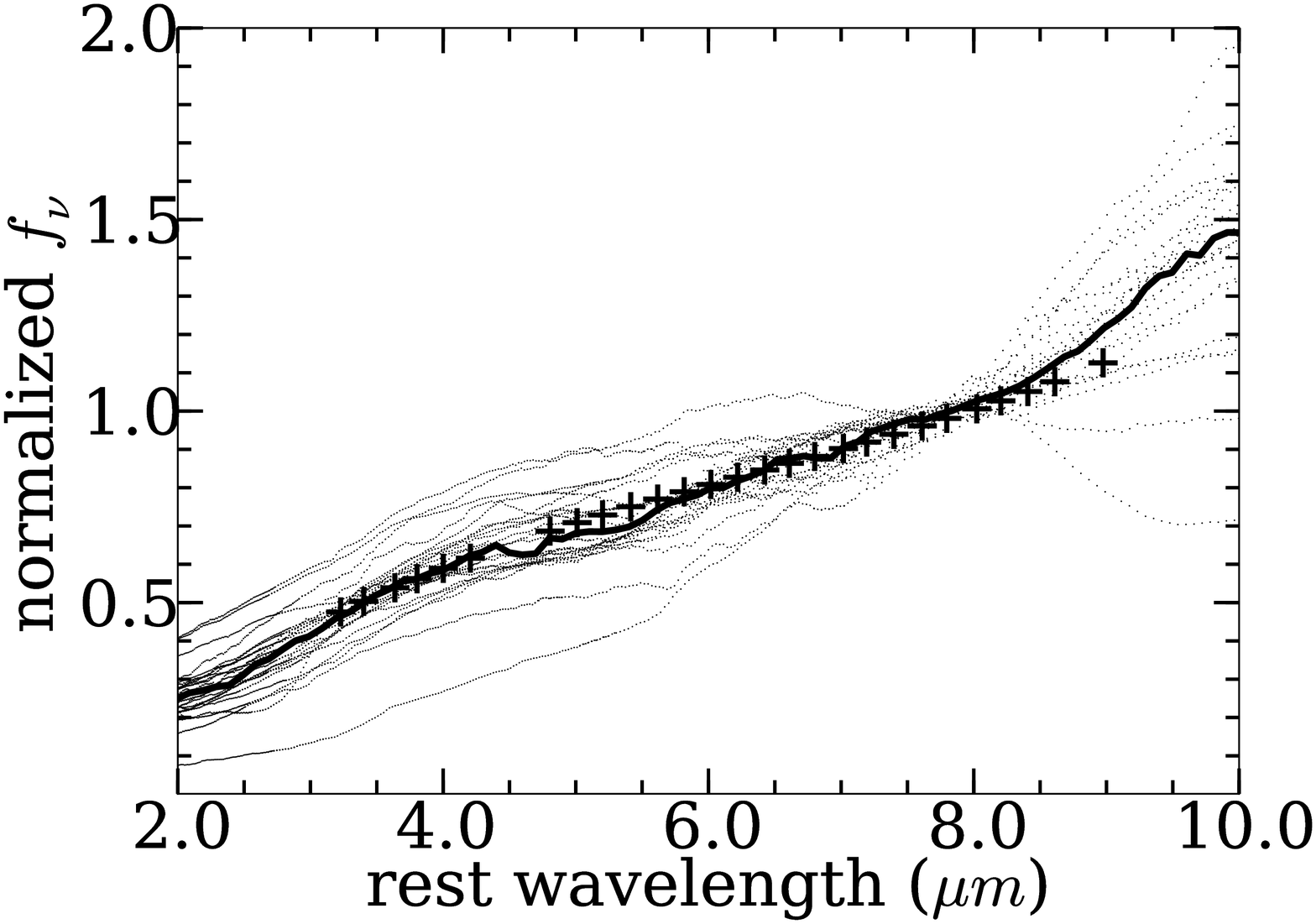}
\caption{Comparison of observed IRS rest frame spectra of SDSS quasars with template used to calculate $f_{\nu}$(7.8 \ums), normalized to $f_{\nu}$(7.8 \ums) = 1 mJy.  Template from \citet{wee12} is shown by crosses and has form $f_{\nu}$(rest frame) = $0.46 + 0.027 \lambda + 0.0038\lambda^{2}+ 0.000148\lambda^{3}$, normalized to 1 mJy at 7.8 \ums.  Dotted curves are individual spectra of SDSS quasars in \citet{deo11}, moved to rest frame according to the SDSS redshift. Spectra are taken from the Cornell Atlas of Spitzer IRS Spectra (CASSIS; Lebouteiller et al. 2011).  Thick solid curve is median of these spectra. } 

\label{template}

\end{figure}

To transform observed WISE $f_{\nu}$(22 \ums) to monochromatic observed flux density $f_{\nu}$(7.8~\ums) at the wavelength corresponding to rest frame 7.8 \ums, the mid-infrared template spectrum from \citet{wee12} is used.  This template is derived using spectra obtained with the $Spitzer$ IRS of type 1 AGN and high redshift SDSS quasars.  It is shown in Figure 1 together with the individual IRS spectra of SDSS quasars at z $\sim$ 2 \citep{deo11} taken from the spectral atlas of Lebouteiller et al. (2011)\footnote{http://cassis.astro.cornell.edu/atlas. CASSIS is a product of the Infrared Science Center at Cornell University.}. For the 22 \um bandpass, the template applies for redshifts 1.4 $\leq$ z $\leq$ 6.3 for the template rest frame 3 \um $<$ $\lambda$ $<$ 9 \um.  The template is restricted to this wavelength range because shorter wavelengths are not dominated by dust emission, and longer wavelengths show large dispersion in spectra because the silicate feature at $\sim$ 10 \um can have widely varying strength, either in emission or absorption.  The low redshift limit of 1.4 set by the template determines the low redshift limit of the SDSS quasars used in our analysis.  

Rest frame luminosity $\nu L_{\nu}$(7.8~\ums) is determined as $\nu L_{\nu}$(7.8~\ums) =  4$\pi D_{L}^{2}[\nu/(1+z)]f_{\nu}$(7.8~\ums), for $\nu$ corresponding to 7.8 \ums, and $D_{L}$ for $\Omega_{M}$=0.27, $\Omega_{\Lambda}$=0.73, and H$_0$ = 74 \kmsMpc \citep{rie11}.  For comparison to $\nu L_{\nu}$(7.8~\ums), rest frame ultraviolet luminosities $\nu L_{\nu}$(0.25~\ums) are determined in section 3, below, with the same cosmological assumptions.

\begin{figure}

\figurenum{2}
\includegraphics[scale= 0.5]{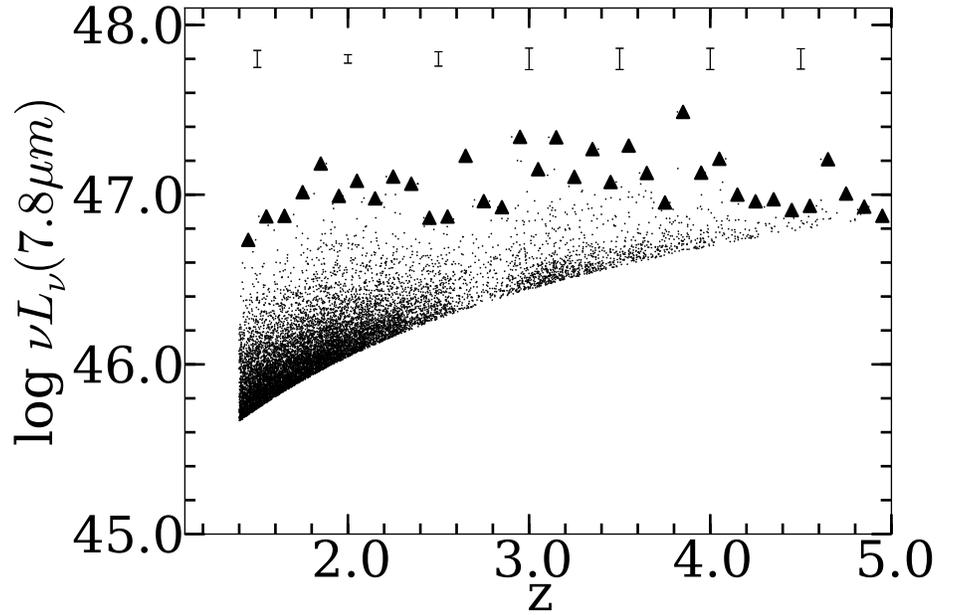} 

\caption{Luminosities $\nu L_{\nu}$(7.8 \ums) in erg s$^{-1}$ compared to redshift for all SDSS quasars with z $>$ 1.4 having measured WISE flux densities at 22 \um (points). Lower envelope arises from detection limit adopted for WISE of 3 mJy at 22 \ums.  Luminosities for SDSS/WISE quasars at rest frame 7.8 \um are determined using spectral template in Figure 1. Vertical error bars are the uncertainties in determining $\nu L_{\nu}$(7.8 \ums) from the observed $f_{\nu}$(22 \ums) arising from dispersion of sources defining the spectral template in Figure 1. Triangles show most luminous sources in each bin of dz = 0.1, and indicate no redshift evolution for the most infrared luminous quasars.}  

\label{78_dist}

\end{figure}

WISE also provides $f_{\nu}$(12 \ums), which reaches to a limit of 0.6 mJy. In principle, use of the 12 \um fluxes could also determine rest frame $f_{\nu}$(7.8 \ums) for redshifts 0.3 $<$ z $<$ 3 using the same spectral template, and this is a potential gain because the $f_{\nu}$(12 \ums) has a 3$\sigma$ upper limit of 0.6 mJy compared to 3 mJy for $f_{\nu}$(22 \ums).  However, comparison of results showed a systematically fainter $f_{\nu}$(7.8 \ums) when determined from $f_{\nu}$(12 \ums) compared to determinations for the same objects using $f_{\nu}$(22 \ums), using a magnitude zero point of 29.0 Jy for WISE 12 \um magnitudes.

This inconsistency led us to check the WISE fluxes using the SDSS quasars with IRS infrared spectra illustrated in Figure 1 by measuring monochromatic flux densities in the IRS spectra at observed wavelengths of 22 \um and 11.5 \um (the effective wavelength of the WISE ``12 \um" filter for most spectral shapes as described in Wright et al. 2010).  This comparison showed agreement between the observed $f_{\nu}$(22 \ums) (IRS/WISE ratio of 0.95 $\pm$ 0.12) but a systematically larger IRS/WISE ratio of 1.25 $\pm$ 0.10 for WISE $f_{\nu}$(12 \ums) compared to IRS $f_{\nu}$(11.5 \ums). The WISE 12 \um filter is very broad, which means that the monochromatic effective wavelength is dependent on spectral shape \citep{wri10}.  This effect may account for most of the discrepancy rather than any systematic offset in flux calibration.   Until the source of this discrepancy is better understood, we do not utilize the 12 \um measures, but the numbers of sources that could be added in different redshift bins using $f_{\nu}$(12 \ums) when $f_{\nu}$(22 \ums) $<$ 3 mJy are summarized in Table 1. 

There are 11959 SDSS quasars detected by WISE at 22 \um out of 31801 SDSS quasars in the relevant redshift range.  After removing known gravitationally lensed quasars \citep{ina12}, the calculated $\nu L_{\nu}$(7.8 \ums) compared to redshift is depicted in Figure~\ref{78_dist} for these quasars detected by WISE. (The 100 most luminous SDSS/WISE quasars are individually tabulated in Weedman et al. 2012). This Figure is the empirical result for the observed infrared dust luminosities of the brightest known quasars as a function of redshift.  No evolution in luminosity is seen. 


\subsection{Tracing Luminous Dust in SDSS Quasars with Luminosity Functions}

The observations in Figure 2 provide the initial illustration that the most luminous quasars in the SDSS/WISE sample when measured with dust luminosity seem uniformly distributed for all z $\ga$ 1.5, with no indication of a particular epoch at higher redshifts at which the most luminous sources are found. We now examine this observed result more quantitatively using the space density of luminous quasars by determining luminosity functions at different z.


The luminosity functions we use are determined as integral functions, expressed as the space density of objects brighter than a given luminosity with the standard ``1/V$_{a}$" technique, using the redshift bins given in Table 1 for the SDSS/WISE quasars.  Using the notation in \citet{mcg13}, for example, $\rho$($>$$L$, z) is the space density of quasars more luminous than $L$ within any redshift bin centered on z, determined as $\rho$($>$$L$, z) = N($>$$L$, z)/V$_{a}$ for V$_{a}$ the co-moving volume of the redshift bin.  For all luminosity functions, $L$ is $\nu L_{\nu}$(7.8 \ums) determined as described above. 

Uncertainties in the luminosity functions arise from $\sqrt{N}$ statistical uncertainties for the number of sources used to determine a value of $\rho$($>$$L$, z), and from uncertainties in the template transformation to rest frame 7.8 \ums.  The redshift bins and number of objects within each bin are given in Table~1 for the SDSS/WISE quasars.  Volumes for each bin are determined for H$_0$ = 74 \kmsMpc, $\Omega_{M}$=0.27 and $\Omega_{\Lambda}$=0.73 using the 9380 deg$^{2}$ sky coverage of the SDSS DR7 quasar catalog.

\clearpage

\begin{deluxetable}{cccc} 
\tablecolumns{4}
\tabletypesize{\footnotesize}

\tablewidth{0pc}
\tablecaption{Redshift Bins for SDSS/WISE Luminosity Functions}

\tablehead{
 \colhead{Redshift Bin} &\colhead{\# of QSO\tablenotemark{a}} & \colhead{\# of QSO\tablenotemark{b}} &\colhead{\# of QSO\tablenotemark{c}}
\\
\colhead{} & \colhead{0.25 \um} &  \colhead{WISE 22 \um} & \colhead{WISE 12 \um} 
}
\startdata

1.4 - 1.5& 3020 &1533 &	1117 \\  
1.5 - 1.6& 3421 &1593 &	1356 \\ 
1.6 - 1.7& 3032 &1378 &	1197 \\
1.7 - 1.8& 3061	&1295 &	1256 \\
1.8 - 1.9& 2771 &1153 &	1175 \\   
1.9 - 2.0& 2354 &1023 &	935 \\ 
2.0 - 2.2& 3379 &1370 &	1464 \\
2.2 - 2.4& 2016 &758 &	897 \\
2.4 - 2.6& 1077 &433 &	443 \\
2.6 - 2.8& 415  &150 &	157 \\
2.8 - 3.0& 880 &213 &	198 \\ 
3.0 - 3.2& 1770 &331 &	0 \\ 
3.2 - 3.4& 1488 &236 &	0 \\ 
3.4 - 3.6& 757 &148 &	0 \\ 
3.6 - 3.8& 874 &132 &	0 \\ 
3.8 - 4.0& 568 &92 &	0 \\
4.0 - 5.0& 918 &121 &	0 \\ 
\enddata

\tablenotetext{a}{Number of quasars in the redshift bin having $f_{\nu}$(0.25 $\mu$m)}
\tablenotetext{b}{Number of quasars in the redshift bin for which the 7.8~\um luminosity is calculated using WISE $f_{\nu}$(22~\ums) $>$ 3 mJy.}
\tablenotetext{c}{Number of quasars in the redshift bin for which the 7.8~\um luminosity could be calculated using WISE $f_{\nu}$(12~\ums) $>$ 0.6 mJy. }

\end{deluxetable}

\begin{figure}

\figurenum{3}
\includegraphics[scale= 0.5]{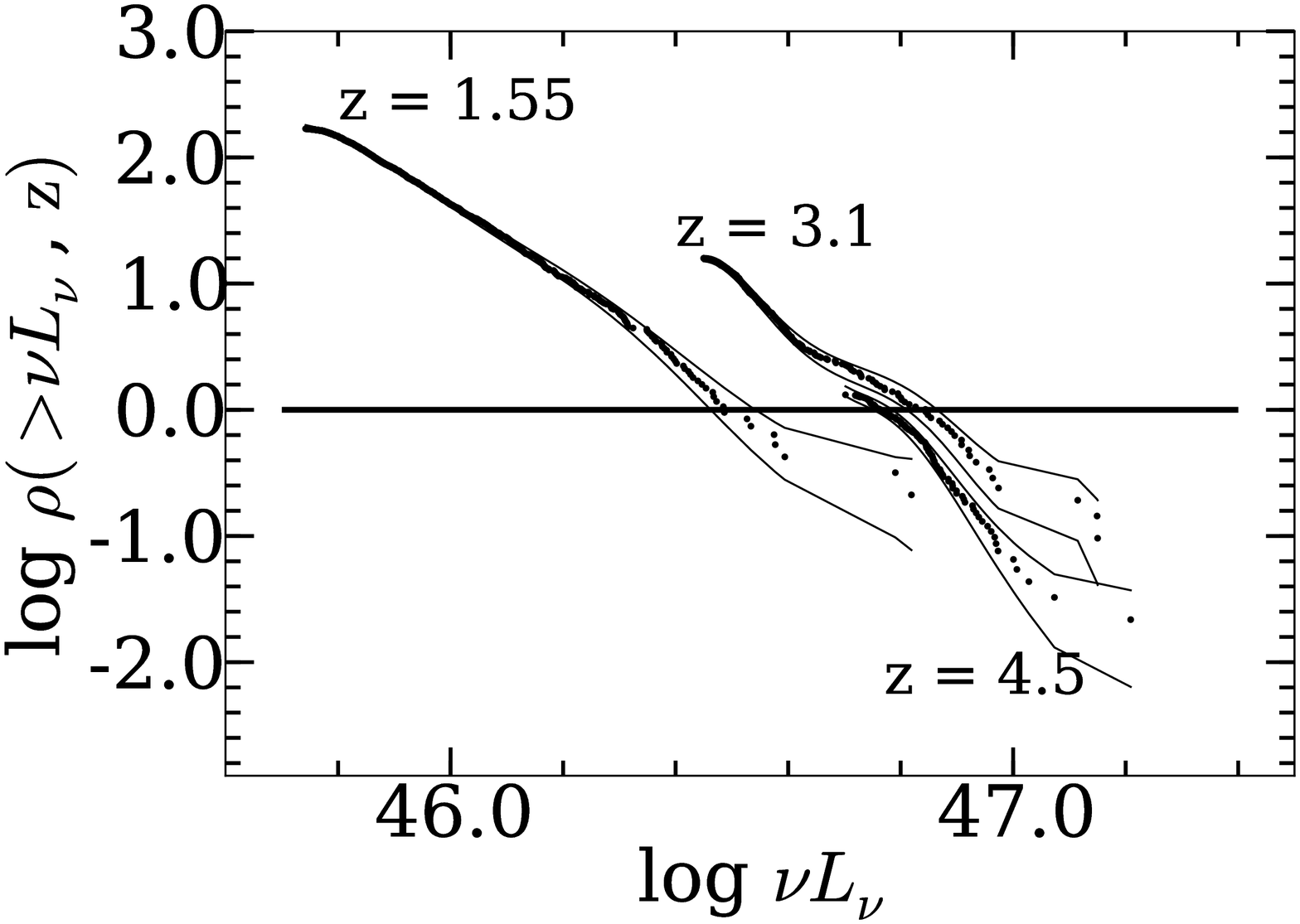}
\caption{Sample luminosity functions for $\nu L_{\nu}$(7.8 \ums) in erg s$^{-1}$  for different redshifts giving density $\rho$($>$$L$, z) in Gpc$^{-3}$.   Upper function is for z = 1.55, center for z = 3.1, and lower for z = 4.5.  All redshift bins for which luminosity functions are determined are given in Table 1.  Upper and lower envelopes for luminosity functions are determined using statistical uncertainties $\sqrt{N}$ for N the number of quasars $>$ $L$ in a particular redshift bin; differences between these envelopes are shown as error bars in Figure 4 below.  Horizontal line shows space density of one quasar Gpc$^{-3}$ more luminous than $\nu L_{\nu}$(7.8 \ums) used for tracing pure luminosity evolution in Figure 4. The faint end of the luminosity functions is determined by the luminosity corresponding to the flux density limit of 3 mJy for WISE $f_{\nu}$(22 \ums). } 

\label{LF_78}

\end{figure}

Figure~\ref{LF_78} shows examples of calculated luminosity functions for $\nu L_{\nu}$(7.8 \ums) within 3 redshift bins. The plots show each luminosity function, as well as polynomial fits corresponding to the upper and lower extremes of the luminosity function in a particular redshift bin that would arise from $\sqrt{N}$ uncertainties for the number of quasars $>$$L$.  Such luminosity functions are determined for all redshift bins in Table 1 using the quasars in each bin having WISE 22 \um detections.  As seen for the examples in Figure 3, the luminosities which can be measured with the WISE detections do not reach deep into the luminosity functions (only a factor of 10 at even the lowest redshift).  This does not affect our analysis because we need only the most luminous sources, but the shape and cutoff of the luminosity functions at the faint ends are determined by observational selection effects so cannot be used for quantitative comparisons.

Evolution of quasar luminosity functions is defined as pure luminosity evolution (for which luminosity functions at all redshifts change only in luminosity but not in space density), as pure density evolution (for which all luminosity functions change only in space density but not in luminosity), or as some combination of the two.  Our objective is to quantify changes with redshift of some high luminosity within the luminosity functions for the SDSS/WISE quasars to measure luminosity evolution for the brightest quasars.  This goal requires a measure of luminosity functions only at the most luminous (bright) end, because we are concerned only with locating the most luminous quasars and determining their space densities.  The characteristic luminosity which we measure is the luminosity $L$ defined by $\rho$($>$$L$, z) for the same space density at all redshifts.  Choosing a space density that includes only the most luminous quasars and tracing the change of this luminosity with redshift defines our measure of luminosity evolution. 


To compare space densities and determine luminosity evolution, a value of $\rho$($>$$L$, z) is required that intersects all luminosity functions at luminosities which are included within the observed luminosity functions at all redshifts.  A density value that satisifies this requirement is found to be one quasar Gpc$^{-3}$ (10$^{-9}$ Mpc$^{-3}$) $>$ $L$.  This value of space density is shown as the horizontal line in Figure 3, which intersects different luminosity functions at different values of $L$ for different redshift bins. Luminosity evolution is measured as the change of $L$ with redshift.  Statistical uncertainties lead to ranges for the luminosities corresponding to this space density at different redshifts.  We calculate these uncertainties as the difference between polynomial fits to the lower and upper luminosity functions arising from smaller and larger values of N $\pm \sqrt{N}$.  The intersection of the horizontal line for $\rho$($>$$L$, z) = 1 Gpc$^{-3}$ with these two fits for each redshift bin, as in the examples of Figure 3, gives the upper and lower values of the luminosity at a given redshift.

The result is shown as crosses in Figure 4 using all of the SDSS quasars for which we have measured $L$ = $\nu L_{\nu}$(7.8 \ums).  The observed luminosity evolution shows a decrease for 2 $<$ z $<$ 3, flattens for z $>$ 3, and remains constant to the highest redshift bin observed (4 $<$ z $<$ 5).  This structure in the luminosity evolution reflects irregularities in the SDSS survey counts as a function of redshift, as illustrated by \citet{ric06} and \citet{sch10}.

\begin{figure}

\figurenum{4}
\includegraphics[scale= 0.5]{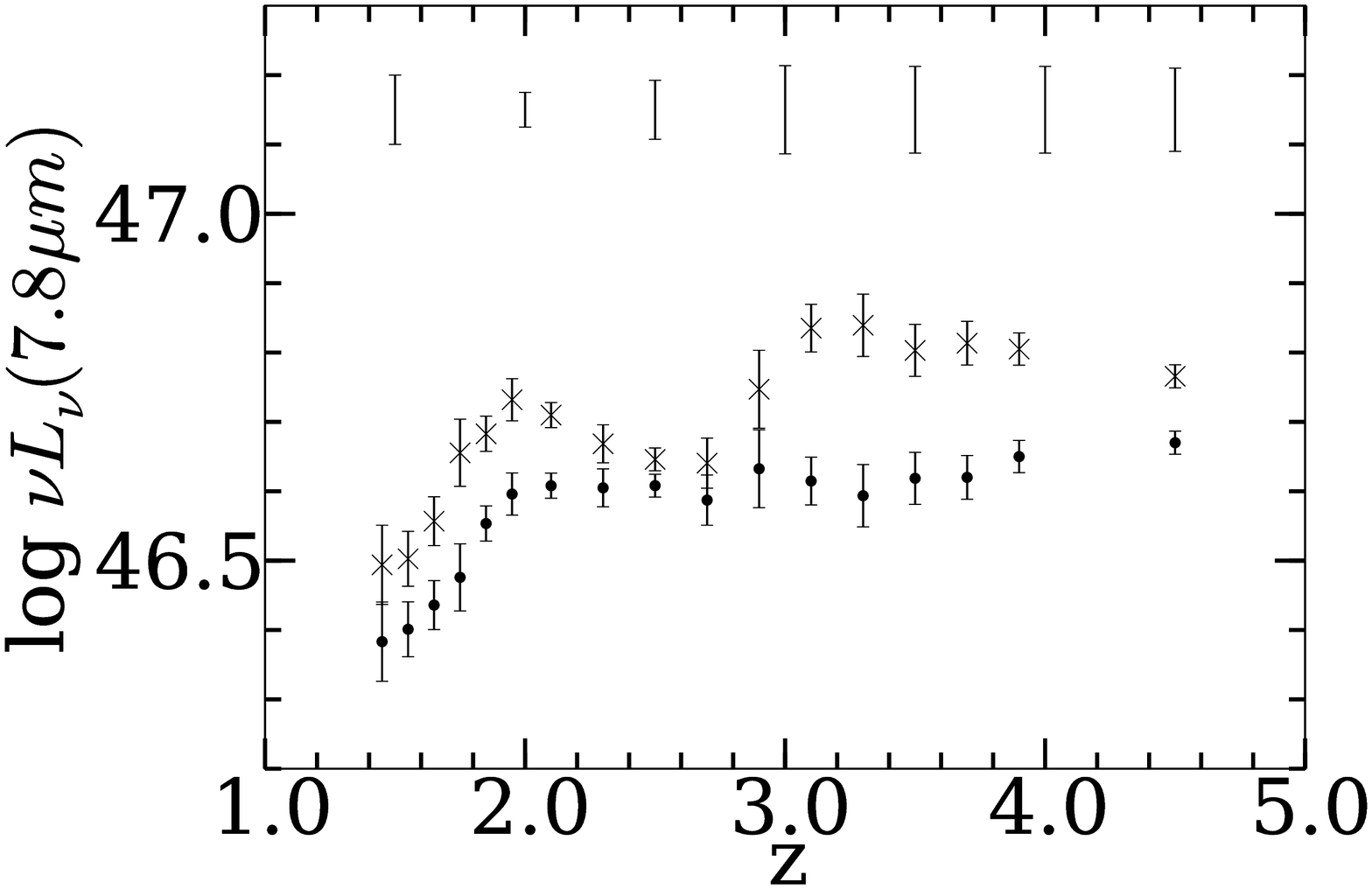}
\caption{Luminosity evolution with redshift for the value of $\nu L_{\nu}$(7.8 \ums) in erg s$^{-1}$ corresponding to one quasar Gpc$^{-3}$ having luminosity $>$ $\nu L_{\nu}$(7.8 \ums) for quasars in the SDSS.  Crosses are results for the full SDSS sample uncorrected for redshift-dependent selection. Corrected distribution using redshift-dependent selection function illustrated in \citet{sch10} based on quasar continuum luminosities is shown by filled circles.   Error bars derive from N $\pm \sqrt{N}$ uncertainties using the upper and lower envelopes for the luminosity functions at a given redshift, as in Figure~\ref{LF_78}.  Uncertainties in rest frame $\nu L_{\nu}$(7.8 \ums) at different redshifts arising from dispersion of sources defining the spectral template in Figure 1 are shown as separate vertical error bars at the top. } 

\label{Levol_78}
 
\end{figure}

The SDSS is carefully defined regarding the photometric selection criteria \citep{ric02,ric06}, but the photometric selection of quasars for some redshifts is incomplete because of confusion with the colors of stars.  In addition, the presence of different emission lines in the broad band filters at different redshifts means that the photometric selection is not based strictly on continuum luminosities.  Fainter continuum sources can be boosted into the sample and quasar selection enhanced if the photometric magnitudes are made brighter by strong emission lines.  These effects result in apparent irregularities in the redshift distribution of SDSS quasars as discussed in \citet{ric06} and \citet{sch10}. The redshift distributions are corrected to illustrate how the redshift distribution would appear if sources of given continuum luminosity were evenly selected at all redshifts, as shown graphically in Figure 8 of Richards et al. and Figure 7 of Schneider et al. To determine corrected luminosity functions at each redshift bin that we utilize in Table 1, we apply the graphical correction of \citet{sch10} to correct all space densities in the observed luminosity functions.  Corrected space densities within any redshift bin are determined by multiplying the observed SDSS quasar numbers by the factor (corrected quasar count/observed quasar count) within that redshift bin from Schneider et al. Figure 7.

After this correction to the luminosity functions, results for the evolution in infrared luminosity for $\rho$($>$$L$, z) = 1 Gpc$^{-3}$ are shown in Figure 4 as filled circles.  For these corrected luminosity functions, there is no maximum luminosity found at any redshift; the distribution flattens and remains constant for all z $>$ 2.  This behavior of luminosity evolution at a defined space density mimics at a slightly lower luminosity the redshift distribution in Figure 2 for the most luminous quasars observed at different redshifts. This result confirms the earlier conclusion from Figure 2 that no distinct epoch has yet been observed when quasars had their maximum infrared luminosity.

These conclusions contrast with the observations and models reviewed in Section 1 that show decreasing ultraviolet luminosities for z $\ga$ 2.5.  This difference in the evolution of ultraviolet luminosity compared to the dust-reradiated infrared luminosity implies a changing dust content of quasars depending on either epoch or luminosity.  In the next section, we compare infrared dust luminosities $\nu L_{\nu}$(7.8 $\mu$m) with observed ultraviolet luminosities $\nu L_{\nu}$(0.25 \ums) to seek any observational indications of changing dust content.

\section{Comparing Ultraviolet and Infrared Luminosities}

The results in the preceding Section find no epoch of peak luminosity for z $<$ 5 when quasars are measured by dust luminosity $\nu L_{\nu}$(7.8 \ums). In this Section, we determine if the fractional dust content of SDSS quasars changes systematically with luminosity or redshift by comparing infrared dust luminosities $\nu L_{\nu}$(7.8 $\mu$m) with observed ultraviolet luminosities $\nu L_{\nu}$(0.25 \ums) for SDSS quasars from \citet{she11}.  This comparison leads to our quantitative definition of unobscured quasars defined by the ultraviolet to infrared luminosity ratios for SDSS quasars.  We also use these ratios to test some expectations of the dusty merger scenario \citep{hop06,nar10}.  

Obscuring dust in a quasar can affect in two ways the ratio $\nu L_{\nu}$(0.25 \ums)/$\nu L_{\nu}$(7.8 \ums), subsequently ``ultraviolet/infrared ratio (UV/IR)".  If the covering fraction for dust clouds surrounding the quasar increases, a larger fraction of the intrinsic ultraviolet luminosity is absorbed and reradiated by dust.  This increases infrared luminosity, thereby decreasing UV/IR, even if the observed ultraviolet luminosity is not diminished by dust extinction.  For constant covering fractions, the observed ultraviolet luminosity can decrease (as in type 2 sources) if dust happens to intervene and lower the observed ultraviolet luminosity because of extinction, which lowers UV/IR without any enhancement of the infrared luminosity.  

The UV/IR ratio, therefore, is a measure of dust content regardless of whether the dust leads to increased IR or decreased UV.  We use the term ``obscuration" to refer to this change in observed dust content, whether arising from increased covering factor for dust and resulting increase in infrared luminosity,  or arising from increased extinction of ultraviolet luminosity by dust in the line of sight.  Either effect of obscuration decreases UV/IR, so examining quasars for systematic change in UV/IR as a function of redshift or luminosity determines if there is a systematic change in dust content.  Our interpretations of systematic changes in UV/IR which follow are applied only to statistical medians for samples of sources; individual objects can deviate from the medians because of various factors such as differences in extinction laws, opacity of obscuring dust clouds, and dust geometry.

\subsection{Observed Ultraviolet/Infrared Luminosity Ratios for SDSS Quasars}

The SDSS quasar data in \citet{she11} give a consistent, monochromatic measure of ultraviolet continuum flux density, which is the $f_{\nu}$(0.25 $\mu$m) at the observed wavelength corresponding to rest frame 0.25 \ums.  The $\nu L_{\nu}$(0.25 \ums) is a measure of observed luminosity from the accretion disk which can be compared to reradiated dust luminosity $\nu L_{\nu}$(7.8 \ums).  

The ultraviolet/infrared ratio for the SDSS/WISE quasars is shown in Figures 5 and 6, compared to both luminosity and to redshift.  These comparisons are restricted to the most luminous quasars, log $\nu L_{\nu}$(0.25 \ums) $>$ 47, and to z $<$ 4 in order to maximize the fraction of 22 \um detections. These luminosity and redshift limits are chosen so that the number of upper limits on $\nu L_{\nu}$(7.8 \ums) for sources undetected by WISE (corresponding to lower limits on the ultraviolet/infrared ratio) is small enough that the median value of the ratio at most redshifts and luminosities is not affected by detection limits. 


\begin{figure}

\figurenum{5}
\includegraphics[scale= 0.5]{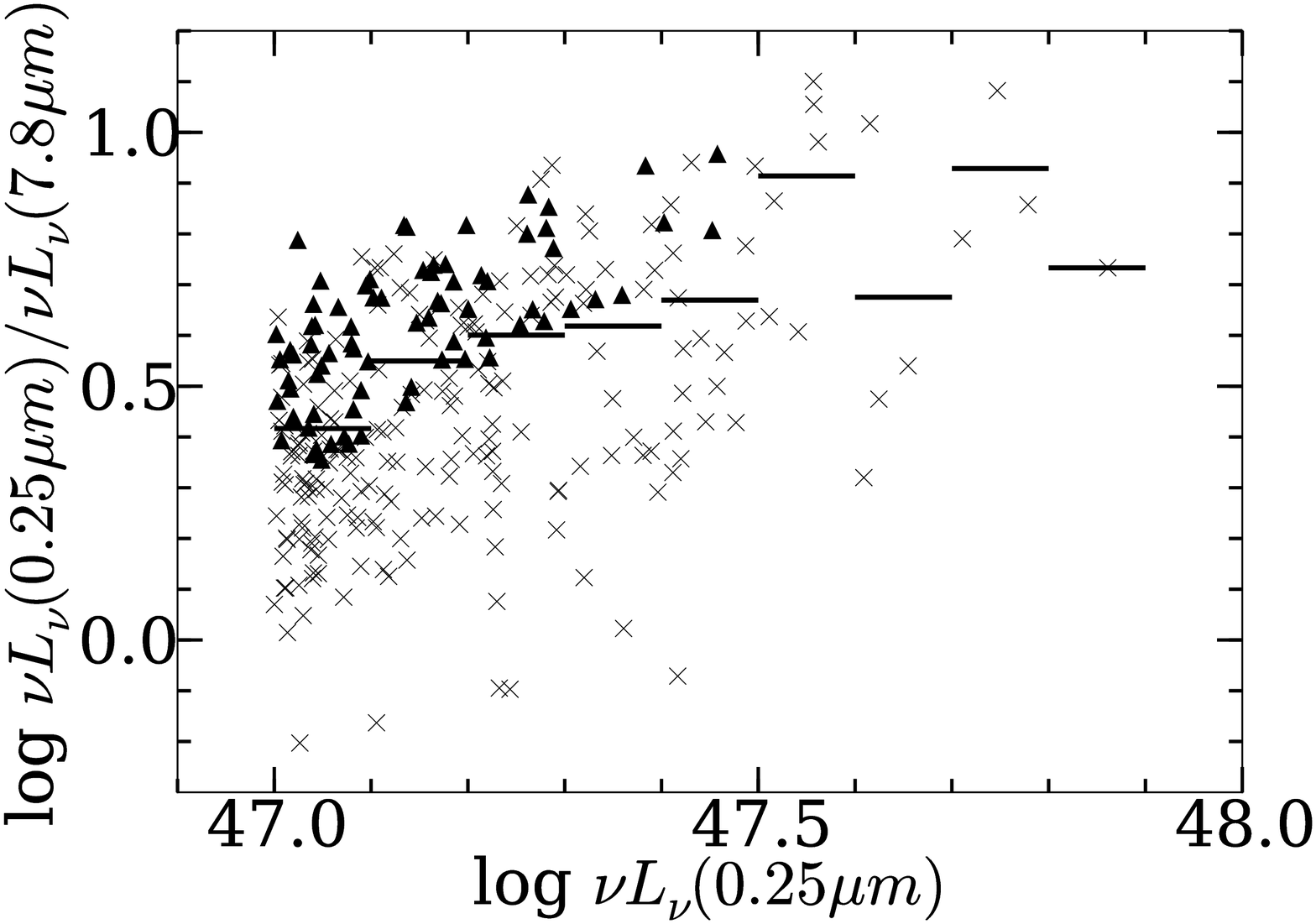}

\caption{Comparison of ratio $\nu L_{\nu}$(0.25 \ums)/$\nu L_{\nu}$(7.8 \ums) with luminosity in erg s$^{-1}$ for all quasars having log $\nu L_{\nu}$(0.25 \ums) $>$ 47 and  1.4 $<$ z $<$ 4. Crosses are SDSS quasars with WISE 22 \um detections ($f_{\nu}$(22 \ums) $>$ 3 mJy) and triangles are quasars with lower limits to ratio arising from $f_{\nu}$(22 \ums) $<$ 3 mJy. Horizontal lines are medians within each range of d[$\nu L_{\nu}$(0.25 \ums)] = 0.1. }

\label{ratio_lum}
\end{figure}

\begin{figure}

\figurenum{6}
\includegraphics[scale= 0.5]{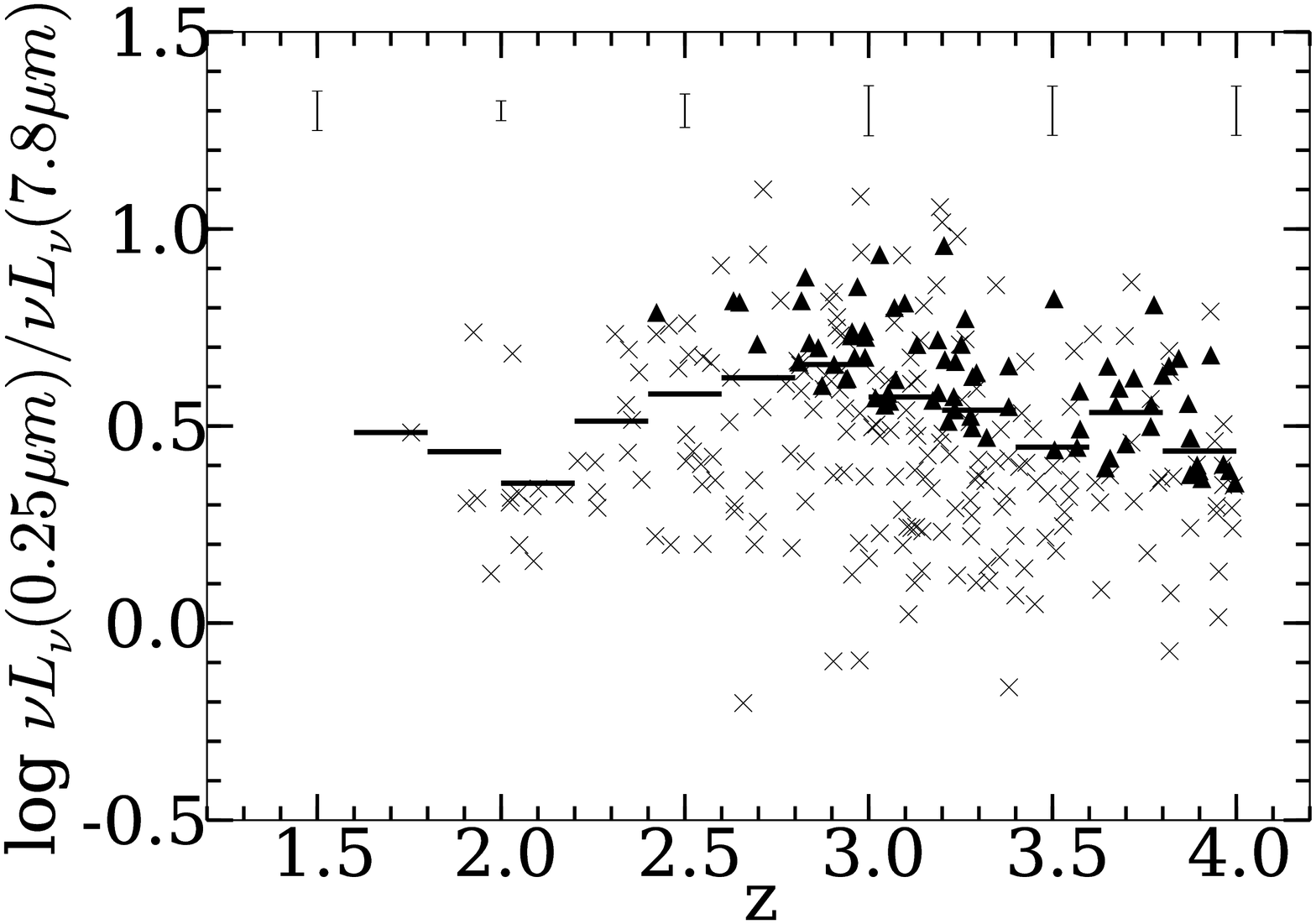}

\caption{Comparison of ratio $\nu L_{\nu}$(0.25 \ums)/$\nu L_{\nu}$(7.8 \ums) with redshift for all quasars having log $\nu L_{\nu}$(0.25 \ums) $>$ 47 and  1.4 $<$ z $<$ 4. Crosses are SDSS quasars with WISE 22 \um detections ($f_{\nu}$(22 \ums) $>$ 3 mJy) and triangles are quasars with lower limits to ratio arising from $f_{\nu}$(22 \ums) $<$ 3 mJy. Horizontal lines are medians within each range of dz = 0.2.  Uncertainties in rest frame $\nu L_{\nu}$(7.8 \ums) at different redshifts arising from dispersion of sources defining the spectral template in Figure 1 are shown as separate vertical error bars at the top.  } 

\label{ratio}
\end{figure}




Because quasars in the SDSS are the brightest known in the rest frame ultraviolet, they must include those quasars having minimal extinction of the ultraviolet.  Consequently, we subsequently refer to these as ``unobscured" quasars, and the ultraviolet/infrared luminosity ratio for such quasars is defined by the results in Figures 5 and 6.  In Section 4, a specific comparison is made between luminosity functions of unobscured and obscured quasars within the redshift interval 1.8 $\le$ z $<$ 2.4.  Figure 6 shows that in this redshift range, the smallest value of log [$\nu L_{\nu}$(0.25 \ums)/$\nu L_{\nu}$(7.8 \ums)] is $\sim$ 0.2 for SDSS/WISE quasars.  

This value is adopted as the limiting ultraviolet/infrared ratio for unobscured quasars; no SDSS/WISE quasars are more obscured than this limiting value even though this limiting value may include some amount of extinction.  Larger values of the ultraviolet/infrared ratio would correspond to higher ultraviolet luminosities, possibly because of less extinction. This limiting value of 0.2 is used in section 4.1 to define the amount of ultraviolet obscuration seen in the sample of obscured quasars used for comparison to the SDSS quasars.

\subsection{Comparison to Dusty Merger Models}

The results in Figures 5 and 6 can test some of the expectations of the dusty merger models for producing luminous quasars and the associated massive galaxies.  These models \citep[e.g.][]{hop06, hop08,nar10} explain the epoch of maximum observed ultraviolet luminosity for quasars (z $\sim$ 2.5) as that time when quasars had maximum accretion rates leading to maximum bolometric luminosities.  The ultraviolet luminosities become visible as the surrounding dust clouds disperse. One expectation from these models, therefore, is that quasars of the highest bolometric luminosity should still have larger covering fractions for dust clouds than cleaner quasars of less luminosity. Observational support for dusty merger models is provided by conclusions that the most luminous quasars with z $<$ 3 do indeed contain the most dust \citep{gli12,uru12}.  

For the quasars with large dust covering fractions to be also those which are most luminous in the ultraviolet, as explained in the dusty merger models, the ultraviolet luminosity must escape between surrounding dust clouds without extinction along the observer's line of sight.  In such cases, a high ultraviolet luminosity should be accompanied by low values of UV/IR, because the large dust covering fraction must enhance the fraction of bolometric luminosity absorbed by the dust and reradiated in the infrared.  


The comparison of UV/IR to ultraviolet luminosity in Figure \ref{ratio_lum} does not show this expected result. The medians as well as the upper and lower envelopes of points in Figure 5 indicate that UV/IR increases by a factor of $\sim$ 3 when the ultraviolet luminosity increases by a factor of 10.  This observational result indicates that dust content is minimized in the most ultraviolet luminous quasars.  This could happen because either the covering fraction of dust clouds decreases as the luminosity increases, thereby reducing the infrared luminosity, or there is less intervening dust producing ultraviolet extinction, thereby increasing the observed ultraviolet luminosity.  Based on IRS spectra observing silicate emission in quasars and type 1 AGN, \citet{mai07} concluded that the explanation is because the covering fraction decreases with increasing ultraviolet luminosity.  

Another expectation arising from the dusty merger models is that the epoch (z $\sim$ 2.5) of maximum observed ultraviolet luminosity when surrounding dust clouds just begin to dissipate should show quasars with larger dust content than at later epochs (lower redshifts) when quasar luminosities have declined and dust content has diminished.  This predicts that UV/IR would be systematically smaller at z $\ga$ 2.5 than for lower redshifts.  This expectation is tested in Figure~\ref{ratio}, comparing the ratio $\nu L_{\nu}$(0.25 \ums)/$\nu L_{\nu}$(7.8 \ums) to redshift.  

No systematic trend in ultraviolet/infrared ratio with redshift is observed in Figure~\ref{ratio} that indicates a decrease in dust content and increase in UV/IR as quasars at z $\la$ 2.5 dissipate the surrounding dust.  The maximum ratio, log $\nu L_{\nu}$(0.25 \ums)/$\nu L_{\nu}$(7.8 \ums) = 0.65, occurs at z = 2.9 rather than at lower redshifts, although this maximum is not significantly different from the median ratio (0.49) within the uncertainties.  The simplest interpretation of this result is that the ultraviolet/infrared ratio is primarily controlled by changes in the ultraviolet extinction along the line of sight instead of by changes in the covering fraction of the dust.  In this case, the maximum UV/IR at z = 2.9 arises because of minimum extinction at that redshift.

In summary, none of the results in Figures 5 and 6 are consistent with explaining the apparent peak in quasar ultraviolet luminosity at z $\sim$ 2.5 as a real peak in bolometric luminosity using the simplified expectations of the dusty merger scenario.  Maximum quasar ultraviolet luminosities which are observed seem to be determined primarily by small changes in extinction rather than by particular stages of mergers which control the accretion rate and covering fraction of surrounding dust clouds.  Our analysis of UV/IR, therefore, does not find any supporting evidence based on this parameter that the epoch of peak intrinsic bolometric luminosity for quasars has been found.

\section{Luminosity Functions for Obscured and Unobscured Quasars at z $\sim$ 2}

The preceding analyses of dust content for SDSS quasars omits any population of quasars so obscured by dust that they are not found in the SDSS.  Yet, dusty quasars are the initial phases of luminous quasars according to the dusty merger scenario, so dusty quasars should dominate quasar samples at the earliest epochs.  Tracing the covering fraction of dust as a function of redshift would measure the transition from early, heavily obscured quasars to later, unobscured quasars. Infrared observations make it possible to discover quasars that were not included in optically derived samples because of heavy dust obscuration (the DOGs; Dey et al. 2008).

Our primary goal in this section is to compare luminosity functions for heavily obscured quasars to luminosity functions for the unobscured quasars in optical surveys.  Surveys of obscured quasars are not yet adequate to make this comparison over a wide redshift range.  Because of the redshift selection effects in the sample of obscured quasars, this determination can most reliably be done at a redshift z $\sim$ 2.  With such a luminosity function as reference, eventual surveys of dusty quasars that reach to higher redshifts can define evolution of the luminosity function at higher redshifts for dust obscured quasars. 

\subsection{Defining Obscured Quasars}

Various studies have revealed a population of obscured AGN, usually considered as type 2 AGN, that is comparable with or exceeds type 1 AGN, perhaps depending on luminosity \citep{wil00,ale03,zak04,mar06,hic07}.  This obscured population has been confirmed photometrically in the infrared using colors from the $Spitzer$ Infrared Array Camera (IRAC; Fazio et al. 2004), sometimes combined with X-ray fluxes, by assuming that those sources with power law continua in the IRAC bands are AGN \citep{brn06,don07,fio08,bus09,mel12}.  

A variety of observing programs using surveys at 24 \um with the $Spitzer$ MIPS instrument \citep{rie04} combined with subsequent IRS spectra were designed to find dust obscured populations of optically faint sources.  As reviewed in Section 1, the obscured population found in this way led to recognition of the DOGs. The DOGs overlap in characteristics with many of the Compton thick, obscured X-ray sources \citep{pol08,fio08,bau10}.

DOGs have redshifts derived from IRS spectra using either strong polycyclic aromatic hydrocarbon (PAH) emission features, with the strongest at 7.7 \ums, or strong silicate absorption centered at 9.7 \ums.   By analogy to similar spectra in low redshift sources having both optical classifications and IRS spectra \citep[e.g.][]{sar11}, the PAH sources are starbursts and the absorbed sources are AGN \citep[e.g.][]{saj07,fel13}.  The quasars among DOGs are defined, therefore, by a spectroscopic classification criterion. 

Optical redshifts and classifications cannot be determined for the most obscured DOGs at high redshift \citep{des08}, and the large resulting uncertainties in estimating the luminosity function of type 2 quasars are summarized by \citet{bro06}, who conclude that the type 2/type 1 ratio has an upper limit of $\sim$ 2 for z $>$ 1.  This agrees with estimates for power-law DOGS which are also X-ray sources, but \citet{don07} derive a ratio of 4:1 using all power law infrared sources.  Comparison with X-ray luminosities to determine Compton thick sources indicates that obscured quasars among the DOGs may outnumber visible quasars by $\sim$ 2:1 \citep{bau10,pol08}.  

There is ambiguity in comparing various studies of DOGs and other obscured sources because there is no consistent definition of an ``obscured" source compared to ``unobscured".  Comparing the optically discovered SDSS/WISE quasars discussed in preceding sections with the $Spitzer$-discovered DOGs allows a quantitative comparison of these two extreme populations in terms of (rest frame) ultraviolet/infrared ratios.  The DOGs used in the analysis which follows have $I$ $>$ 24, approximately 5 magnitudes fainter than the faintest SDSS/WISE quasars, but have comparable infrared flux density of a few mJy at 24 \ums.  
 
For the obscured quasars summarized below, the value of $\nu L_{\nu}$(0.25 \ums) cannot be determined precisely because rest frame ultraviolet flux densities are measured only with broad band $R$ and $I$ filters; which one is closer to rest frame 0.25 \um depends on redshift as these filters have effective wavelengths of $\sim$ 0.65 \um and 0.80 \um.  The magnitudes of the brightest obscured quasars are $\sim$ 24 in either filter, and this magnitude would correspond to 0.77 or 0.61 $\mu$Jy for $R$ or $I$.  Taking the average of these flux densities as representing the optically brightest obscured quasar and comparing to the faintest $f_{\nu}$(7.8 \ums) in Table 1 ($\sim$ 1.5 mJy) yields a limiting log [$\nu L_{\nu}$(0.25 \ums)/$\nu L_{\nu}$(7.8 \ums)] $<$ -1.8; no obscured quasar is less obscured than this ratio.   

In section 3.1, it was determined that no SDSS/WISE quasar is more obscured than having log [$\nu L_{\nu}$(0.25 \ums)/$\nu L_{\nu}$(7.8 \ums)] $<$ 0.2.  Comparing unobscured with obscured quasars between these two samples, therefore, represents populations which differ by at least a factor of 100 in the ultraviolet/infrared ratio.  This difference is explained as arising from dust extinction which affects the obscured quasars and defines our definition of obscured compared to unobscured quasars.  The quantitative definition of obscuring dust used in our further discussion, therefore, means dust which reduces the observed ultraviolet luminosity by at least a factor of 100 compared to what would be observed in an unobscured quasar. 

\subsection{The Sample of Obscured Quasars}

Search criteria leading to the DOG samples were defined with various parameters, so all of the different samples are not easily compared regarding selection effects.  As summarized in \citet{saj12} and \citet{mel12}, for example, the optical limits for selection differ among the various surveys.  To make a well defined comparison to the optically selected SDSS/WISE quasars, we use a DOG sample defined only by optical and 24 \um flux limits, $f_{\nu}$(24 \ums) $>$ 1 mJy and $I$ $>$ 24. This sample \citep{hou05, wee06} includes DOGS within the NOAO Deep Wide Field Survey (NDWFS) in Bo{\"o}tes (8.2 deg$^{2}$; Jannuzi and Dey 1999).  For these Bo{\"o}tes DOGs, IRS spectra are available to enable direct measures of $\nu L_{\nu}$(7.8~\ums) and spectroscopic classification as obscured quasar using the silicate absorption feature.  The full sample is given in Table 2 and provides the set of optically obscured quasars for comparison to the much brighter SDSS quasars, which have 19 $<$ $i$ $<$ 20 mag.

We used the CASSIS database of IRS spectra \citep{leb11} to remeasure the spectra of Bo{\"o}tes DOGs. Compared to previously published IRS results, the CASSIS spectra have substantial improvement in signal to noise (S/N) of observed IRS spectra.  The S/N of extracted IRS spectra depends primarily on the precision of background subtraction for faint sources with source intensities only a few percent of the background. The improvements within CASSIS include ``optimal" extraction (weighting of individual pixels according to their fraction of source illumination using empirically determined point spread functions; Lebouteiller et al. 2010) and applying a quantitative S/N criteria to choose the best background subtraction.  

Use of the improved CASSIS spectra allowed redshift determinations for some previously unmeasured sources, improved the starburst/AGN classification, and improved the $\nu L_{\nu}$(7.8~\ums) luminosity measures. The CASSIS spectra are especially valuable in improving S/N at the longest wavelengths of the IRS for which sensitivity and S/N are least ($\ga$ 30 \ums).  This resulted in the discovery of 7 new sources in Table 2 for which redshifts were not determined in previous analyses. 

All Bo{\"o}tes spectra classified as AGN with silicate absorption and having z $>$ 1.5 are summarized in Table 2 with the newly measured redshifts and $f_{\nu}$(7.8~\ums). The average spectrum of all sources with 1.8 $\le$ z $<$ 2.4 used for determining the luminosity function is shown in Figure 7, illustrating the characteristic deep silicate absorption and weak PAH emission. In accordance with previous classifications, the formal criterion that equivalent width (EW) of the 6.2 \um PAH feature be EW(6.2 \ums) $<$ 0.1 \um is used to define a source as an AGN, but most sources included in Table 2 have no measurable 6.2 \um feature that exceeds the spectral noise, and the upper limits are significantly smaller than 0.1 \ums.

\begin{figure}

\figurenum{7}
\includegraphics[scale= 0.5]{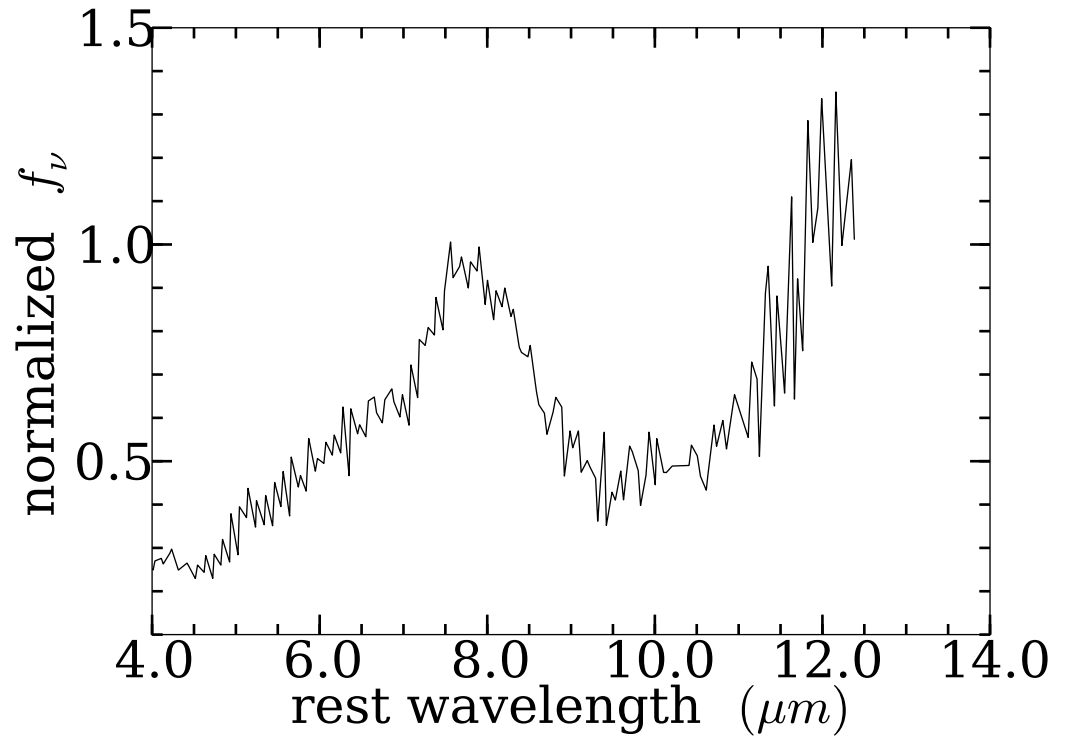}

\caption{Average rest frame spectrum of Bo{\"o}tes obscured quasars in Table 2 having 1.8 $\le$ z $<$ 2.4, normalized to $f_{\nu}$(7.8 \ums) = 1 mJy. The peak at 7.8 \um in the rest frame spectrum of these sources is used to measure $\nu L_{\nu}$(7.8~\ums), and this peak causes selection of such sources to be optimal near z $\sim$ 2 where the spectral peak falls in the $Spitzer$ 24 \um survey filter.}

\end{figure}

For flux density $f_{\nu}$(7.8 $\mu$m), the major source of uncertainty is the statistical scatter among the data points at the spectral peak.  We use as a measure of $f_{\nu}$(7.8 $\mu$m) the median flux density for all points in the spectrum with rest frame 7.7 \um $<$ $\lambda$ $<$ 7.9 \um.  This corresponds to the spectral peak but only includes a single IRS resolution element at typical redshifts.  There are two pixels per resolution element.  The CASSIS extractions combine observations from the two independent nods with separate wavelength calibrations to yield two data points per pixel.  Spectra used for our measurements are not smoothed.  To estimate the uncertainty arising from using so few data points, we also measured the median flux density for rest frame 7.2 \um $<$ $\lambda$ $<$ 8.4 \um and compared the two results.  After accounting for the expected offset because the wider band includes points below the peak, the one sigma scatter about the median ratio of the two measures is 10$\%$.  This is taken as the measurement uncertainty of the results arising from statistical noise.

The luminosities and redshifts of the sources in Table 2 are shown in Figure 8.   The restricted range of redshifts for the absorbed quasars arises from the selection effects in the $Spitzer$ discovery technique.  These sources were initially found in 24 \um surveys so selection favors sources for which the continuum peak at rest frame 7.8 \um shown in Figure 7 falls within the 24 \um photometry band.  This effect leads to the concentration of sources centered at z $\sim$ 2.1.  To compromise between a narrow redshift range and adequate sources for meaningful statistics, space densities are determined for the Bo{\"o}tes quasars which have 1.8 $\le$ z $<$ 2.4.  This provides the optimal high redshift interval for which we can determine a luminosity function for obscured quasars having spectroscopic redshifts and classifications.

\begin{figure}

\figurenum{8}
\includegraphics[scale= 0.5]{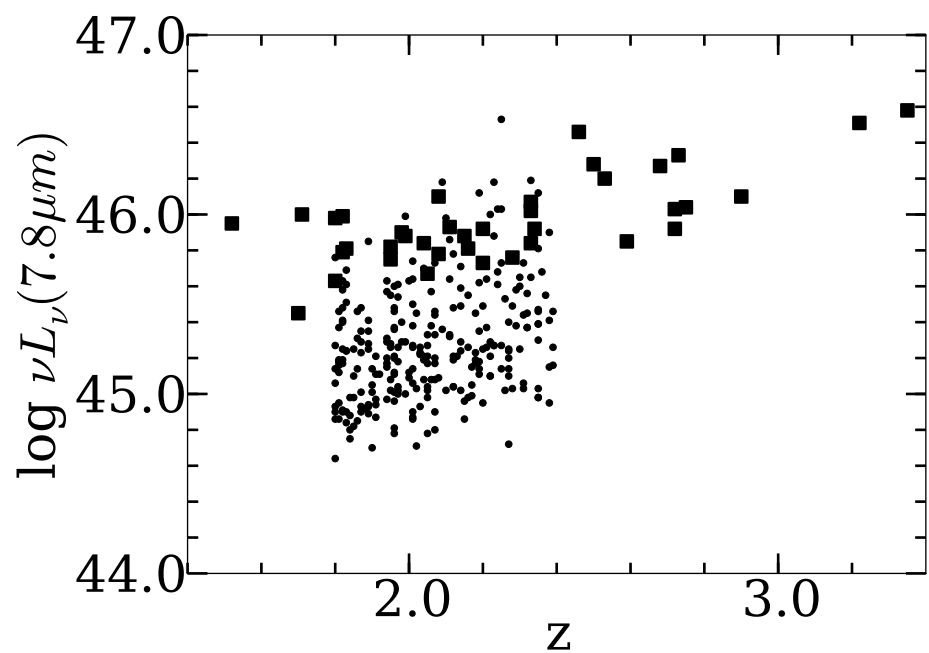} 

\caption{Luminosities $\nu L_{\nu}$(7.8 \ums) in erg s$^{-1}$ compared to redshift for all Bo{\"o}tes obscured quasars in Table 2 (squares) and for all AGES quasars with 1.8 $\le$ z $<$ 2.4 (circles).  The AGES quasars arise from the same survey area as the Bo{\"o}tes obscured quasars.  Obscured quasars used for the luminosity function in Section 4.3 are the subset with 1.8 $\le$ z $<$ 2.4.}

\end{figure}

There is only a small luminosity range at which luminosities for these obscured quasars overlap with the SDSS/WISE quasar luminosities shown in Figure 2. The small survey area of Bo{\"o}tes does not encompass statistically rare, very luminous sources found in SDSS/WISE, whereas the brighter 3 mJy flux limit of SDSS/WISE does not reach the faint luminosities of the 1 mJy Bo{\"o}tes sources.  

The AGN and Galaxy Evolution Survey (AGES; Kochanek et al. 2012) for optical quasars provides the crucial overlap. AGES covers the same Bo{\"o}tes survey area as the obscured quasars using optically discovered quasars which also have measured infrared luminosities.  Luminosities for AGES quasars reach limits approximately ten times fainter than SDSS quasars in both optical and infrared luminosities because the AGES luminosities derive from a $Spitzer$ 24 \um survey reaching 0.3 mJy for sources having $R$ $<$ 23.5.  The AGES $\nu L_{\nu}$(7.8 $\mu$m) reach fainter than the Bo{\"o}tes obscured quasars and also overlap the luminosities of some SDSS/WISE quasars while probing much deeper into the infrared luminosity function for optically discovered quasars. 

Luminosities $\nu L_{\nu}$(7.8 $\mu$m) are measured for the 265 AGES quasars with 1.8 $\le$ z $<$ 2.4 using the tabulated 24 \um flux densities and optical redshifts in \citet{koc12} combined with the spectral template in Figure 1.  At the redshifts being used for this luminosity function, the observed frame 24 \um corresponds to rest frame 7.1 \um $<$ $\lambda$ $<$ 8.6 \ums, so the uncertainties in transforming to rest frame 7.8 \um are small. Infrared luminosities of these AGES quasars are also shown in Figure 8 and compared to the obscured quasars in Table 2 from the same Bo{\"o}tes area.  Because the two surveys include the same area and both are measured photometrically at 24 \ums, comparisons for the most luminous sources of each type are direct, with no statistical uncertainties arising from comparing different survey volumes.

\clearpage

\begin{deluxetable}{lcccccc} 
\tablecolumns{7}
\tabletypesize{\footnotesize}
\tablewidth{0pc}
\tablecaption{ Luminosities and Redshifts for Obscured Bo{\"o}tes Quasars }
\tablehead{
 \colhead{Source} &\colhead{name and coordinates} &\colhead{z\tablenotemark{a}} & \colhead{$f_{\nu}$(7.8 $\mu$m)\tablenotemark{b}}& \colhead{ $\nu L_{\nu}$(7.8 \ums)\tablenotemark{c}}& \colhead{$R$\tablenotemark{d}} & \colhead{Ref.\tablenotemark{e}} 
\\
\colhead{} & \colhead{J2000} &  \colhead{} & \colhead{mJy} & \colhead{log erg s$^{-1}$} & \colhead{Vega Mag.} & \colhead{} 
}
\startdata

1 & SST24 142538.23+351855.1\tablenotemark{f} & 2.28 & 1.3 & 45.76 & $>$25.4 & 1\\    
2 & SST24 142611.35+351217.9 & 1.82	 &	 2.1 & 45.79 & $>$26 & 2 \\
3 & SST24 142622.01+345249.2 &1.98	 &	2.3 & 45.90 & 24.5 & 3\\
4 &  SST24 142648.90+332927.2 &1.82	 &	 3.3 & 45.99 & 24.2 & 3	\\
5 & SST24 142653.23+330220.7 &1.80	 &	1.5 & 45.63 & 25.7  & 3	\\
6 & SST24 142745.88+342209.0 &3.35	 &	4.5 & 46.58 & 24.2 & 2 \\
7 & SST24 142804.12+332135.2 & 2.16	 &	1.6 & 45.81 & 25.5 & 3\\	
8 &  SST24 142924.83+353320.3	& 2.05	 &	1.3 & 45.67 & 25.6 & 1\\	 
9 & SST24 142931.36+321828.2 &2.33	 &	1.5 & 45.84 &  $>$26 & \nodata \\
10 &  SST24 142958.33+322615.4 & 2.34	 &	1.8 & 45.92 & 25.5 & 1\\	 
11 & SST24 143001.91+334538.4 &2.46	 &	 5.8 	& 46.46 & 24.8  & 1\\
12 &   SST24 143004.77+340929.9 &3.22	 &	4.1 & 46.51 & 23.9 & 2\\ 
13 & SST24 143025.74+342957.3 &2.73	 &	3.6 & 46.33 & 23.9 & 3	\\ 
14 & SST24 143026.04+331516.3 & 1.83 & 2.2  & 45.81 & 24.3 & 5 \\
15 & SST24 143028.52+343221.3 &2.15	 &	1.9 & 45.88	 & 24.5 & 4\\
16 & SST24 143048.34+322532.2\tablenotemark{g} & 1.71 & 3.8  & 46.00 & 23.1 & 5 \\ 
17 & SST24 143109.78+342802.7 &2.20	 &	 1.3 & 45.73 & $>$25.5  & 3\\
18 & SST24 143135.29+325456.4 &1.52	 &	4.3 & 45.95 & 23.9 & 3\\
19 & SST24 143251.89+333536.8 & 1.70 & 1.1 & 45.45 & $>$25.5 & 3\\
20 &  SST24 143253.39+334844.3 &2.90	 &	1.9 & 46.10 & $>$26 & 2 \\
21 &  SST24 143301.49+342341.5 \tablenotemark{g} & 2.22   & 2.4  & 46.00 & 22.8 & 5\\ 
22 & SST24 143312.70+342011.0 &2.11	 &	 2.2  & 45.93 & 24.2	& 4\\
23 & SST24 143318.59+332127.0  & 2.72 	 &	 1.4 	& 45.92 & $>$26 & 2\\ 
24 & SST24 143358.07+332607.7  & 1.95 & 1.7 & 45.75 & $>$25.9 & 3\\
25 & SST24 143410.96+331732.9\tablenotemark{g} & 2.73 & 1.8 & 46.03 & 22.2 & 6 \\ 
26 & SST24 143447.70+330230.6  & 1.99 	 &	 2.2  & 45.88 & 26.0& 3	\\  
27 & SST24 143504.12+354743.2 &2.08	 &	 1.6 	& 45.78 & $>$25.8 & 3\\
28 & SST24 143508.49+334739.8  & 2.08  &	 3.4 	& 46.10 & 24.1 & 3\\
29 & SST24 143520.75+340418.2 & 2.20	 &	 2.0 	 & 45.92 & $>$25.1 & 1\\
30 &  SST24 143523.99+330706.8  &2.59	 &	 1.3 	& 45.85 & 25.0 & 1\\
31 & SST24 143539.34+334159.1	&2.50	 &	 3.7 	 & 46.28 & $>$25.5 & 1\\
32 & SST24 143545.11+342831.4 &2.53	 &	 3.0 	& 46.20 & $>$25.2 & 3\\
33 & SST24 143644.22+350627.4 & 1.80  &	 3.3	 & 45.98 & 24.3 & 1\\
34 & SST24 143725.23+341502.4 & 2.04 & 1.9  & 45.84 & $>$25.4 & 3\\
35 & SST24 143807.92+341612.4 & 2.33 & 2.6  & 46.07 & 25.5 & 2 \\ 
36 & SST24 143808.34+341015.6 & 2.33 & 2.3  & 46.02 & 24.6 & 3\\
37 &  SST24 143834.92+343839.3\tablenotemark{g} &2.68	 &	 3.2  & 46.27 & 22.44 & 5 \\ 
\enddata
\tablenotetext{a}{Redshift z measured from fitting median template of AGN with silicate absorption from \citet{sar11}.  Uncertainty in z is $\pm$ 0.08 as determined from scatter in new measures of z compared to original measures using other templates or independent spectra.}
\tablenotetext{b}{Peak flux density at 7.8 \um determined by median of all points in spectrum between 7.7 \um and 7.9 \ums.  Relative uncertainty among sources is $\pm$ 10$\%$ percent.  Absolute uncertainty of CASSIS flux calibration applied to all sources is below $\pm$ 3$\%$. }
\tablenotetext{c}{Rest frame luminosity $\nu L_{\nu}$(7.8 $\mu$m) in erg s$^{-1}$ determined as $\nu L_{\nu}$(7.8 $\mu$m) =  4$\pi$D$_{L}$$^{2}$[$\nu$/(1+z)]$f_{\nu}$(7.8 $\mu$m), for $\nu$ corresponding to 7.8 \ums , taking luminosity distances from \citet{wri06}:  http://www.astro.ucla.edu/~wright/CosmoCalc.html, for H$_0$ = 74 \kmsMpc, $\Omega_{M}$=0.27 and $\Omega_{\Lambda}$=0.73. (Log [$\nu L_{\nu}$(7.8 $\mu$m)(\ldot)] = log [$\nu L_{\nu}$(7.8 $\mu$m)(erg s$^{-1}$)] - 33.59.)}
\tablenotetext{d}{Vega $R$ magnitudes from references in final column or from \citet{bus09} and \citet{mel12}.}
\tablenotetext{e}{Notes give reference for original publication of source; 1 =  \citet{hou05}, 2 =  \citet{wee06} (redshifts were not previously determined for these sources), 3 = \citet{bus09}, 4 = \citet{brn07}, 5 = \citet{brn08}, 6 = \citet{dey05}.}
\tablenotetext{f}{Source previously listed as starburst in \citet{wee08} but changed to absorbed AGN based on new extraction.  }
\tablenotetext{g}{Source selected by different criterion than uniform selection described in text of $f_{\nu}$(24 $\mu$m) $>$ 1 mJy and $I$ $>$ 24, so this source is not used in determination of luminosity function at z $\sim$ 2.1.}
\end{deluxetable}

\clearpage

\subsection{Comparisons of Unobscured and Obscured Infrared Luminosity Functions}

Luminosity functions are calculated as described in Section 2 for Bo{\"o}tes obscured quasars and AGES optical quasars, using the volume contained within 1.8 $\le$ z $<$ 2.4 for the Bo{\"o}tes survey area.  A correction for incompleteness of the Bo{\"o}tes obscured quasars is applied.  As described in \citet{wee06}, the Bo{\"o}tes quasars in Table 2 arise from IRS observations of 54\% of the complete sample of 24 \um sources in the 8.2 deg$^{2}$ Bo{\"o}tes survey field having $f_{\nu}$(24 \ums) $>$ 1 mJy and $I$ $>$ 24 mag.  The space densities determined from sources in Table 1 are increased, therefore, by a factor of 1.9 to produce the luminosity function of obscured sources for comparison to the SDSS/WISE and AGES quasars. 

For this redshift range 1.8 $<$ z $<$ 2.4, the selection function for SDSS quasars does not require any incompleteness corrections to the observed space densities.  According to the selection function in Table 1 of \citet{ric06}, the mean fraction of actual quasars within these redshifts which are found according to the SDSS photometric selection is 0.95.  Such a small correction would change the space densities by an amount smaller than the symbol size in Figures 9 and 10 which compare luminosity functions.   

\begin{figure}

\figurenum{9}
\includegraphics[scale= 0.5]{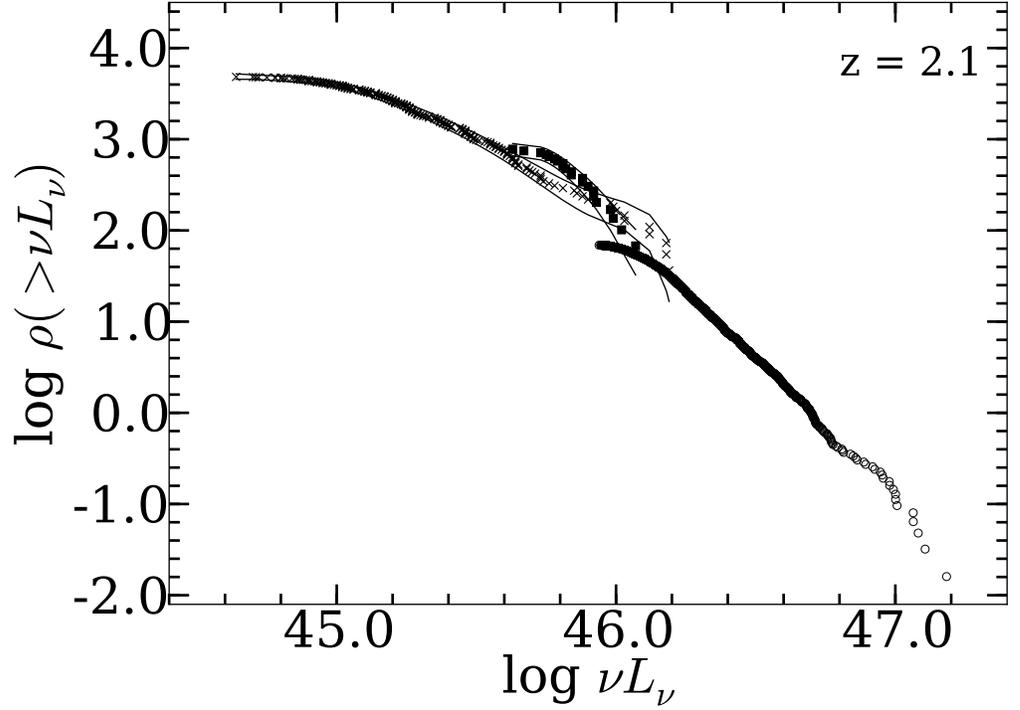}

\caption{Comparison of luminosity functions for luminosities $\nu L_{\nu}$(7.8 \ums) in erg s$^{-1}$ within 1.8 $\le$ z $<$ 2.4 for SDSS/WISE quasars (circles), AGES quasars (crosses) and Bo{\"o}tes obscured quasars (squares).  Space densities have units Gpc$^{-3}$. Upper and lower envelopes shown as thin curves for Bo{\"o}tes and AGES luminosity functions are determined from statistical uncertainties $\sqrt{N}$ for N the number of quasars $>$ $L$ in this redshift interval.  Uncertainties in the comparisons of $\nu L_{\nu}$(7.8 \ums) arising from the template in Figure 1 are smaller than the symbols, because the redshifts used are close to the value when WISE 22 \um (for SDSS) or $Spitzer$ 24 \um (for AGES) corresponds to rest frame 7.8 \ums. Plot demonstrates that the luminosity function based on dust emission has been determined over a luminosity range of about 400 for optically discovered quasars.}

\label{fulllumfuncs}
\end{figure}

Comparison of the full luminosity functions at z = 2.1 for Bo{\"o}tes obscured quasars, AGES quasars, and SDSS/WISE quasars is shown in Figure~\ref{fulllumfuncs}. This comprehensive luminosity function covers a factor of 400 in infrared luminosity $\nu L_{\nu}$(7.8 \ums), primarily for the optically discovered quasars.  The obscured, infrared-discovered quasars fill only a factor of 4 in luminosity.  Comparing luminosity functions of unobscured and obscured quasars can be done only within this restricted luminosity range.

\begin{figure}

\figurenum{10}
\includegraphics[scale= 0.5]{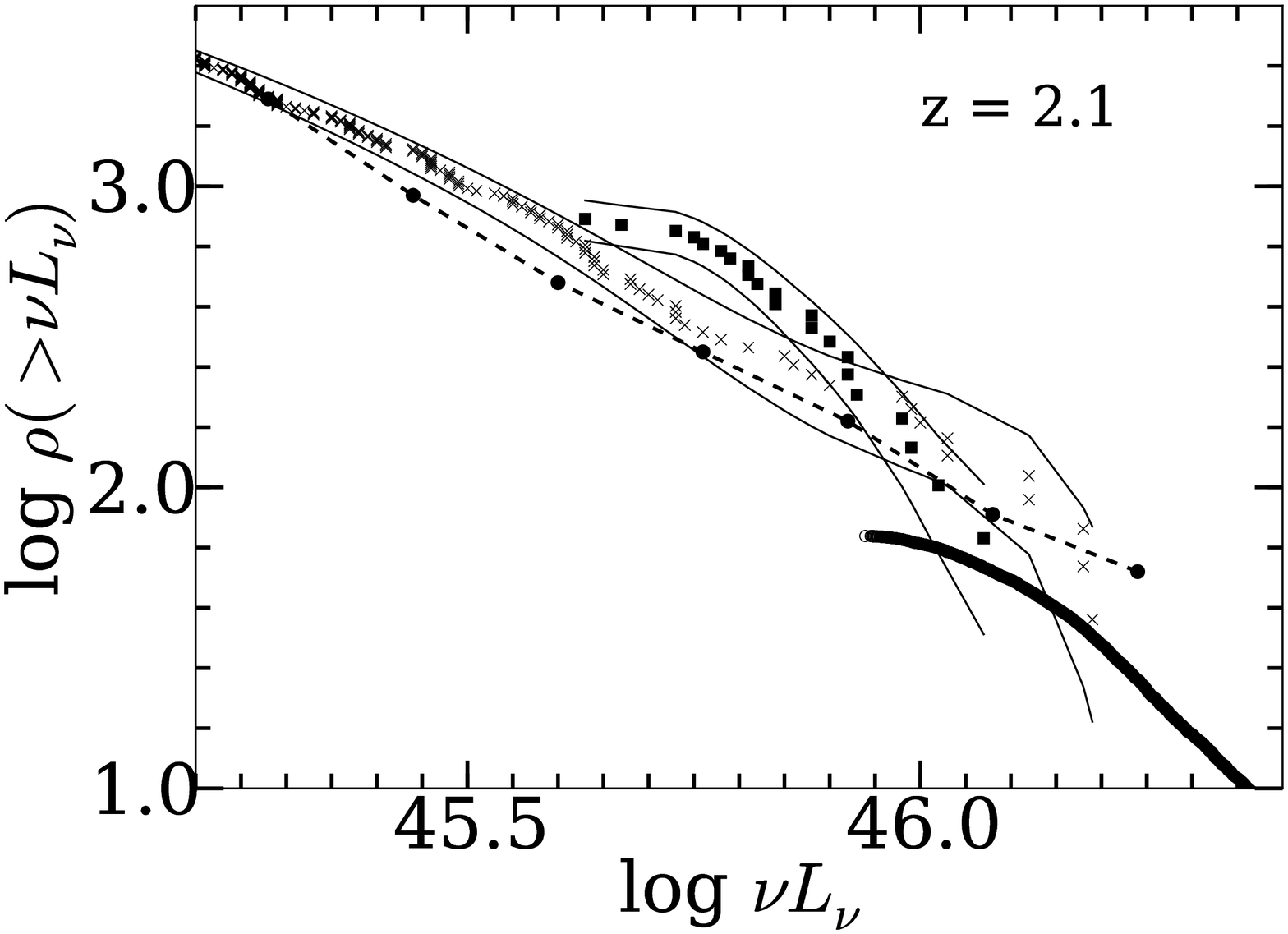}

\caption{Enlarged comparison of luminosity functions from Figure 9 for luminosities $\nu L_{\nu}$(7.8 \ums) in erg s$^{-1}$ within 1.8 $\le$ z $<$ 2.4 for SDSS/WISE quasars (thickest curve), AGES quasars (crosses) and Bo{\"o}tes obscured quasars (squares) discovered with $Spitzer$.  Space densities have units Gpc$^{-3}$. Upper and lower envelopes shown as thin curves for Bo{\"o}tes and AGES luminosity functions are determined from statistical uncertainties $\sqrt{N}$ for N the number of quasars $>$ $L$ in this redshift interval.  Dashed line connects points from $\nu L_{\nu}$(8 \ums) luminosity function of \citet{bro06} for smaller AGES sample within 1.5 $<$ z $<$ 2.5. Plot demonstrates that luminosity functions for optically discovered quasars in SDSS and AGES samples are the same within statistical uncertainties as the luminosity function for obscured quasars within the limited range of overlapping luminosities.}

\label{smalllumfuncs}
\end{figure}

An enlarged portion of the luminosity functions is shown in Figure~\ref{smalllumfuncs} to judge better the overlap between obscured and unobscured luminosity functions. Within the statistical uncertainties, the three luminosity functions overlap.  Perfect overlap would mean that the SDSS and AGES criteria select the same quasars in terms of ultraviolet/infrared luminosity ratios, except that AGES reaches fainter luminosities.  Overlap also means that dust reradiated luminosity functions are the same for optically visible SDSS and AGES quasars compared to the optically obscured Bo{\"o}tes sources.  This result means that the space density of optically bright, unobscured SDSS and AGES quasars is the same at z $\sim$ 2.1 as the optically faint, obscured Bo{\"o}tes quasars.  If the densities were exactly equal for obscured compared to unobscured quasars, the covering fraction of obscuring dust (defined above as dust absorbing at least 5 magnitudes of ultraviolet luminosity) would be 50\%.  Using a similar sample of DOGs observed in the X-ray, \citet{bau10} determined that the obscured fraction defined by X-ray absorption of $N_{H}$ $>$ 10$^{22}$ is $\sim$ 60\%.

The prior study most similar to ours was the effort by \citet{bro06} to determine infrared luminosity functions for the AGES quasars using a smaller sample of these sources reaching only to $f_{\nu}$(24 \ums) $>$ 1 mJy.  This luminosity function was based on AB absolute magnitudes (AB in their notation) derived from $Spitzer$ 24 \um photometry to derive rest frame $\nu L_{\nu}$(8 \ums).  Using their stated tranformation that AB of -28.5 corresponds to log $\nu L_{\nu}$ = 45.61 together with their tabulated luminosity function for 1.5 $<$ z $<$ 2.5 (their Table 3), the Brown et al. result is also shown in Figure 10 (adjusted to the same cosmological parameters). 

Results are consistent, indicating agreement for transforming to rest frame luminosity and also indicating that averaging luminosity functions over slightly different volumes does not significantly change the result.  We note that Brown et al. also measured pure luminosity evolution for these quasars and parameterized the best fit for this luminosity evolution as log $L(z)$ = log $L(z=0)$ + 1.15z - 0.34z$^{2}$ + 0.03z$^{3}$.  This parameterization gives a peak infrared luminosity at z = 2.56.  This result was determined using only 11 quasars with z $>$ 3, however, so it is not inconsistent with the more extensive SDSS results for luminosity evolution above in Section 2 that find no peak luminosity. 



There are some differences that appear in the comparison of luminosity functions which should be noted even though they are statistically marginal. The bright end of the AGES luminosity function is flatter than the bright end of the obscured quasar luminosity function.  If this trend is real, it means that obscured and unobscured quasars differ by some parameter other than simply orientation of the obscuring dust clouds relative to the observer.  The luminosity functions overlap for log $\nu L_{\nu}$(7.8 \ums) $>$ 45.9, but obscured quasars exceed unobscured by a factor of two in space density at log $\nu L_{\nu}$(7.8 \ums) = 45.75.  Taking literally these results would imply that dust content decreases as luminosity increases, the same trend as noted in Section 3.  


A crucial objective for the future is to extend the luminosity function for obscured quasars in Figures 9 and 10 to the brighter luminosities reached by the SDSS quasars so that evolution can be measured for luminous, dusty quasars at higher redshifts.  If the infrared luminosity function for obscured quasars extends to brighter luminosities with the same slope as for SDSS/WISE quasars, obscured quasars should be found in all sky surveys with infrared fluxes comparable to those of the brightest SDSS/WISE quasars.  These can have $f_{\nu}$(22~\ums) up to 30 mJy at z = 2, falling to 3 mJy by z = 5 (scaling from the 3 mJy limit illustrated in Figure 2).

Infrared luminous, optically faint quasars have been found in the WISE survey using mid-infrared photometric color criteria \citep{eis12}, but sources analogous to most of the obscured Bo{\"o}tes DOGs would be too faint for optical redshifts \citep{des08}, and IRS spectra can no longer be obtained for redshift determinations.  Assembling candidate sources for the most luminous obscured quasars from the WISE survey is highly important but also challenging because it requires optical imaging to $R$ or $I$ $\sim$ 26 mag. for each potential WISE obscured quasar.

\section{Summary and Conclusions}

Infrared luminosities $\nu L_{\nu}$(7.8 \ums) arising from dust reradiation are determined for all optically discovered SDSS quasars with z $>$ 1.4 detected at 22 \um with WISE, using an empirical spectral template to transform to rest frame 7.8 \ums.  Infrared luminosity $\nu L_{\nu}$(7.8 \ums) does not show a luminosity maximum at any redshift, reaching a plateau for z $\ga$ 3 with maximum luminosity $\nu L_{\nu}$(7.8 \ums) $\sim$ 10$^{47}$ erg s$^{-1}$ (Figure 2).  Luminosity functions show one quasar Gpc$^{-3}$ more luminous than 10$^{46.6}$ erg s$^{-1}$ for all 2 $<$ z $<$ 5 (Figure 4). From these results, we conclude that the epoch when quasars first reached their maximum luminosity (and the associated epoch of most massive galaxy formation) has not yet been identified at any redshift below 5.  

Comparisons are made between ultraviolet luminosities $\nu L_{\nu}$(0.25 \ums) and infrared luminosities as functions of luminosity and of epoch (Figures 5 and 6).  The quasars most luminous in the ultraviolet have the largest values of $\nu L_{\nu}$(0.25 \ums)/$\nu L_{\nu}$(7.8 \ums)(Figure 5).  This implies that the most ultraviolet luminous quasars are those with the smallest dust content and least extinction of the ultraviolet. No trend or change in ultraviolet/infrared ratio is found within redshifts 1.5 $<$ z $<$ 4 that identifies an epoch where dust content changes systematically, as would be expected if ultraviolet luminosity first appears when surrounding dust clouds dissipate.  

The infrared luminosity functions for SDSS/WISE quasars and for the optically discovered quasars of the AGES survey are determined at z $\sim$ 2.1 and compared to the infrared luminosity function of optically obscured quasars ($I$ $>$ 24) discovered with the $Spitzer$ IRS (Figures 9 and 10).  Based on the observed ultraviolet/infrared luminosity ratios for obscured quasars compared to the ratios for SDSS quasars, obscured quasars are quantitatively defined as having more than five magnitudes of ultraviolet extinction.  Both the unobscured and the obscured quasars at z $\sim$ 2.1 show similar luminosity functions with indications of slightly greater space densities for obscured quasars, indicating a covering fraction for obscuring dust of at least 50\% at this redshift.  

All of these results lead to the conclusion that the apparent maximum in quasar ultraviolet luminosity at z $\sim$ 2.5 is not a measure of the actual peak in intrinsic quasar bolometric luminosity but is probably the result of diminished extinction of the ultraviolet compared to other epochs.  To find the real epoch of maximum quasar luminosity, it is necessary to extend infrared searches for obscured quasars to redshifts z $\ge$ 5.  At such redshifts, the brightest dusty quasars in the sky should have $f_{\nu}$($\sim$ 22~\ums) $\la$ 3 mJy.

\acknowledgments

This publication makes use of data products from the Wide-field Infrared Survey Explorer, which is a joint project of the University of California, Los Angeles, and the Jet Propulsion Laboratory/California Institute of Technology, funded by the National Aeronautics and Space Administration.

We also acknowledge data products from the SDSS. Funding for the SDSS and SDSS-II has been provided by the Alfred P. Sloan Foundation, the Participating Institutions, the National Science Foundation, the U.S. Department of Energy, the National Aeronautics and Space Administration, the Japanese Monbukagakusho, the Max Planck Society, and the Higher Education Funding Council for England. The SDSS Web site is http://www.sdss.org/. The SDSS is managed by the Astrophysical Research Consortium (ARC) for the Participating Institutions. The participating institutions are the American Museum of Natural History, Astrophysical Institute of Potsdam, University of Basel, Cambridge University, Case Western Reserve University, University of Chicago, Drexel University, Fermilab, the Institute for Advanced Study, the Japan Participation Group, Johns Hopkins University, the Joint Institute for Nuclear Astrophysics, the Kavli Institute for Particle Astrophysics and Cosmology, the Korean Scientist Group, the Chinese Academy of Sciences (LAMOST), Los Alamos National Laboratory, the Max-Planck-Institute for Astronomy (MPIA), the Max-Planck-Institute for Astrophysics (MPA), New Mexico State University, Ohio State University, University of Pittsburgh, University of Portsmouth, Princeton University, the United States Naval Observatory, and the University of Washington.

\end{document}